\documentclass[preprint,aps,keywords]{revtex4}
\usepackage[dvips]{graphicx}
\usepackage{array}

\begin{document}
\title{The spatial distribution of dark-matter-annihilation originated gamma-ray line signal}

\author{Tong-Suo Lu $^{1,2}$, Tie-Kuang Dong $^{1}$, Jian Wu $^{1}$}
\address{$^{1}$Key Laboratory of Dark Matter and Space Astronomy,
Purple Mountain Observatory, Chinese Academy of Sciences, Nanjing 210008, China}
\address{$^{2}$Graduate University of Chinese Academy of Sciences, Beijing, 100012, China}

\begin{abstract}
The GeV$-$TeV $\gamma-$ray line signal is the smoking gun
signature of the dark matter annihilation or decay. The detection of
such a signal is one of the main targets of some space-based
telescopes, including Fermi-LAT and the upcoming CALET, DAMPE and
Gamma-400. An important feature of the dark-matter-annihilation
originated $\gamma-$ray line photons is their concentration at the
center of the Galaxy. So far no reliable $\gamma-$ray line has been
detected by Fermi-LAT and the upper limits on the cross section of
annihilation into $\gamma-$rays have been reported. We use these
upper limits to estimate the ``maximal" number of $\gamma-$ray line
photons detectable for Fermi-LAT, DAMPE and Gamma-400 and then
investigate the spatial distribution of these photons. We show that
usually the center of the distribution region will be offset from
the  Galactic centre (Sgr A$^{\star}$) due to the limited
statistics. Such a result is almost independent of the dark matter
distribution models and renders the reconstruction of the dark
matter distribution with the $\gamma-$ray line signal very
challenging for the foreseeable space-based detectors.
\end{abstract}

\maketitle

Dark matter is a kind of special matter necessary
to interpret gravitational effects observed in very large scale
structures that cannot be accounted for by the amount of
observed/normal matter (Jungman et al. 1996; Bertone et al. 2005;
Hooper et al. 2008). Among various candidates, the leading one is
the so-called weakly interacting massive particles (WIMPs), which
may annihilate with each other or decay and then produce particle
pairs such as photons, electrons and positrons and so on(Jungman et
al. 1996; Bertone et al. 2005; Hooper et al. 2008). GeV$-$TeV
$\gamma$-ray line is of extreme interest in search for the signal of
dark matter (DM) annihilation or decay since no other known physical
processes can give rise to similar signal. That's why the detection
of such a signal is one of the main targets of some space-based
telescopes, including Fermi-LAT and the upcoming CALorimetric
Electron Telescope (CALET) (http://calet.phys.lsu.edu),
DArk Matter Particle Explorer (DAMPE) and Gamma-400
(http://gamma400.lebedev.ru/indexeng.html). After analyzing
the publicly available Fermi-LAT $\gamma$-ray data, Bringmann et al.
(2012) and Weniger (2012) found possible evidence for a
monochromatic $\gamma$-ray line with energy $\sim130$ GeV (see also
Tempel et al. 2012 and Su \& Finkbeiler 2012). Such a signal can be
interpreted by $\sim 130$ GeV DM annihilation, that's why it has
attracted wide attentions (Bringmann \& Weniger 2012; Feng et al.
2013; Yang et al. 2013). The offset $\sim 220$ pc (1.5$^0$) of the
center of the most prominent signal region from the Galactic center
Sgr A$^{\star}$ identified by Tempel et al. (2012) and Su \&
Finkbeiler (2012) has been widely taken as a puzzle. This is because
the dark matter distribution is usually assumed to be centered at
Sgr A$^{\star}$, so is the expected dark matter annihilation signal.
Since that such a 130 GeV $\gamma$-ray line signal consists of only
$\sim 14$ photons, the ¡°imperfect¡± consistency of these photons
with the expected dark matter distribution can be just due to the
limited statistics (Yang et al. 2012). However, no firm evidence for
the 130 GeV $\gamma-$ray line emission has been found by the
Fermi-LAT collaboration (Ackermann et al. 2013). Such a negative
result is a bit disappointed. Nevertheless, people are still keen on
detecting $\gamma-$ray line signal, the smoking gun signature of
dark matter annhilation/decay. For example, CALET, DAMPE and
Gamma-400, three upcoming spaced-based telescopes with an energy
resolution $\sim 1-2\%$ above 100 GeV, may contribute significantly
to the $\gamma-$ray line search (Li \& Yuan 2012; Bergstrom et al.
2012). In this work we estimate the morphology of the potential line
signal that is detectable for current and upcoming space-based
$\gamma-$ray detectors. Following Yang et al. (2012) we carry out
the Monte Carlo simulation of the arrival direction of the photons
produced by the annihilation of dark matter particles. The
dependence of the expected ``imperfect morphology" on the dark
matter distribution models is also examined.

   \begin{figure}
   \centering
   \includegraphics[width=6.5cm]{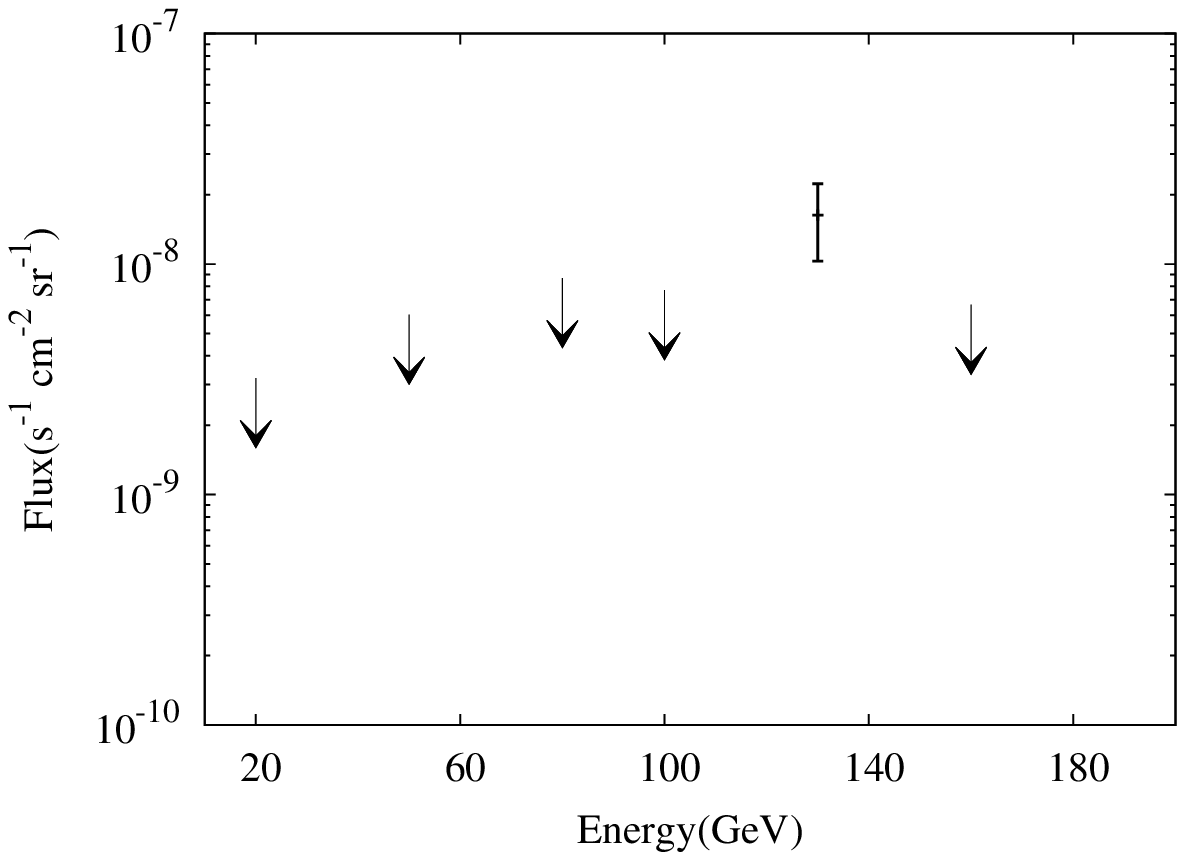}
   \includegraphics[width=7.0cm]{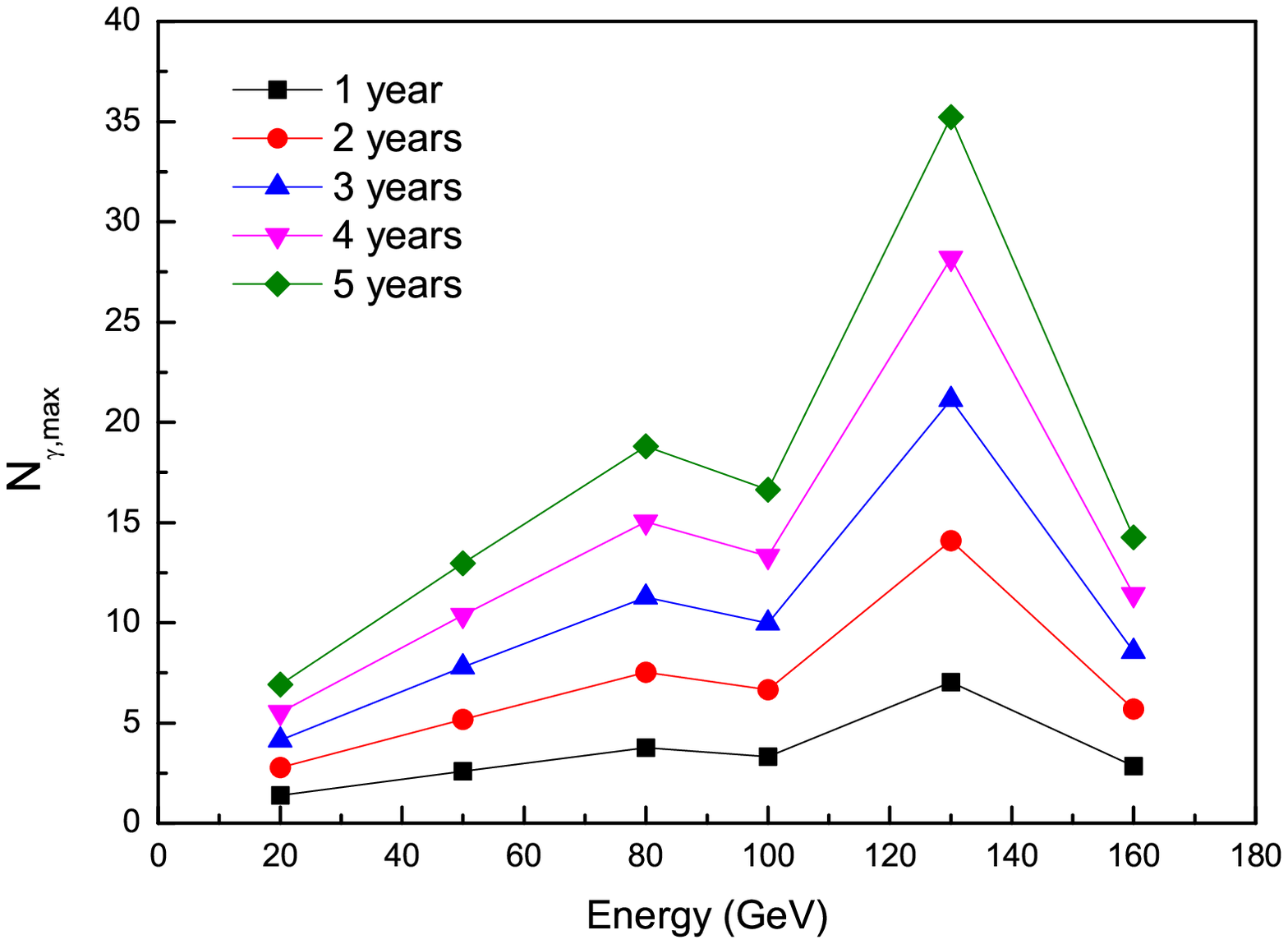}
   \caption{ Left: The upper limits of fluxes of photons with energy $E_{\gamma}$=20, 50, 80, 100, 160 GeV,
   the flux of photons with $E_{\gamma}$=130 GeV is also shown with a local significance of $\sim 3 \sigma$.
   Right: The maximal number of $\gamma-$ray line photons with different energies detectable for Fermi-LAT-like detector
   by different durations of observation. }
   \label{Fig1}
   \end{figure}

For such purposes we analyze the public data of Fermi-LAT to estimate the upper limit on the number of $\gamma-$ray line photons. The standard LAT analysis
software(v9r27p1) (http://fermi.gsfc.nasa.gov/ssc) is adopted. The data set is in the time interval from 4 August 2008 to 18 April 2012, with energies between 20
and 200 GeV.  We take the ULTRACLEAN
dataset to avoid the contamination from the charged
particles. In order to
reduce the effect of the Earth albedo background, time intervals
when the Earth was appreciably in field-of-view, in particular
when parts of the region of interest (ROI) were observed at zenith
angles $> 100^ \circ$, were excluded.  Our spectral analysis was carried out based on the P7v6 version of
post-launch instrument response functions.

We use the
un-binned analysis method which is similar to the one described in
Ackermann et al. (2013) to search the line signal at different energies. The likelihood is described as
\begin{equation}
{\cal L}=\prod_i f S(E_i)+(1-f) B(E_i),
\end{equation}
where $S(E_i)$ and $B(E_i)$, both normalized to $1$, represent the signal and background
function, respectively; the signal fraction $f$ has been set to be in the
range $[-1, 1]$ ($[0, 1]$) for line signal search (getting upper
limits), and $i$ runs over
all the photons (Ackermann et al. 2013).  $B(E_i)$ takes the form
\begin{equation}
B(E_i) \sim E_i^{-\Gamma} \epsilon(E_i),
\end{equation}
where $\epsilon(E_i)$ is the exposure generated by the
\emph{gtexpcube2} routine. Note that $S(E_i)$ is derived by convolving the
energy dispersion
function (http://fermi.gsfc.nasa.gov/ssc/data/analysis
/documentation/Cicerone/Cicerone\_LAT\_IRFs/IRF\_E\_dispersion.html)
and exposure. We take Pyminuit (http://code.google.com/p/pyminuit).
to obtain the maximum of the likelihood and adopt the MINOS
asymmetric error at the level $\Delta ln {\cal L} =1.35$ to
 get upper limit corresponding to a coverage probability of
$95\%$ (see also Yang et al. 2013).

We chose $E_\gamma=20,~50,~80,~100,~130,~160~{\rm GeV}$ and the
ROI is the inner 3 degrees of the galactic center.
For 130 GeV we got a signal with a flux of $1.63 \pm 0.6 \times
10^{-8} {\rm cm^{-2} s^{-1} sr ^{-1}}$ with a local significance of
$\approx 3\sigma$, consistent with Yang et al. (2013). At other
energies we got the upper limits, the results are shown in Fig.1
(left). In Fig.1 (right) we show the detectable photons for
Fermi-LAT like detector as functions of the observation time and the
photon energy. One can see that the potential $\gamma-$ray line photons are very limited.

In this work we consider several dark matter density profiles that have been
widely adopted in the literature, including the so-called
generalized Navarro-Frenk-White profile (NFW, Navarro et al. 1997),
Einasto profile (Einasto et al. 1965), and the Isothermal profile
(Bahcall et al. 1980).
The generalized NFW DM density profile (Navarro et al. 1997) reads
\begin{equation}
\rho(r)=\frac{\rho_{\rm s}}{\left(\frac{r}
{r_{\rm s}}\right)^{\alpha}\left(1+\frac{r}{r_{\rm s}}\right)^{3-\alpha}},
\end{equation}
where $r_{\rm s}\approx 20$ kpc and $\rho_{s}\approx 0.11$
GeV cm$^{-3}$. $\alpha \sim 1$ is found in many numerical simulations. However, in the presence of
baryonic compression $\alpha$ can be high up to $\sim 1.7$. In this work we take $\alpha \sim 1.7$
since such an extreme cuspy distribution can ``narrow" the concentration region of the signal photons.

The Einasto DM density profile (Einasto et al. 1965) reads
\begin{equation}
\rho(r)=\rho_{\rm s,ein}\exp\left(-\frac{2}{a}\left[\left(\frac{r}
{r_{\rm s,ein}}\right)^{a}-1\right]\right),
\end{equation}
where $a=0.17$, $r_{\rm s,ein}\approx 20$ kpc and $\rho_{\rm s,ein}\approx 0.06$
GeV cm$^{-3}$.

The Isothermal DM density profile (Bahcall et al. 1980) reads
\begin{equation}
\rho(r)=\frac{\rho_{\rm s,iso}}{1+\left(\frac{r}{r_{\rm s,iso}}\right)^2},
\end{equation}
where $r_{\rm s,iso}\approx 5.0$ kpc and $\rho_{\rm s,iso}\approx 1.16~{\rm
GeV~ cm^{-3}}$.

The possibility of detecting one $\gamma-$ray in the direction $(\ell,~b)$ is
proportional to the $J$-factor
\begin{equation}
J=\frac{1}{\rho_\odot^{2} R_\odot}\int {\rm d}s \rho^2(r(s)),
\end{equation}
where $\rho_\odot=0.3 $ GeV cm$^{-3}$ is the dark matter density in
the local area, $s$ is
the line of sight distance, $r_\odot \simeq 8.5$
kpc is the distance from the Sun to the Galactic center, $r=(s^{2}+r_\odot^{2}-2sr_\odot \cos \ell \cos
b)^{1/2}$ is the Galactocentric distance, and $(l,~b)$ are the
Galactic longitude and latitude, respectively.

\begin{figure}
\centering
\includegraphics[width=10.0cm, angle=0]{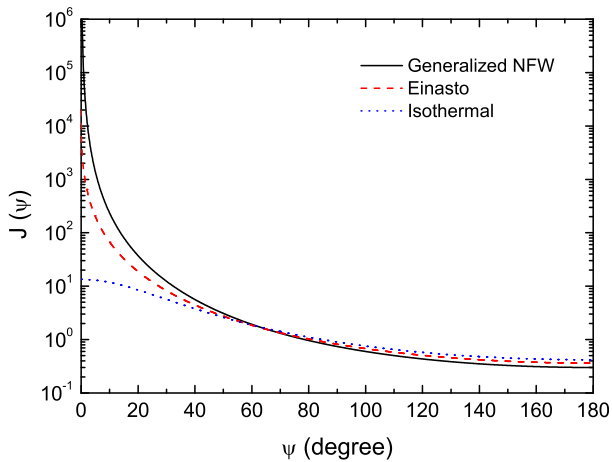}
\caption{The $J$-factor corresponding to the generalized NFW ($\alpha=1.7$),
Einasto, and Isothermal DM density distribution, respectively.}
\label{Fig2}
\end{figure}

Fig.2 shows the $J$-factor for different DM density
distributions changing over the observed direction (the angle
between the line of sight and the direction of galactic center). If
the signal photons are lot enough, their distribution as a function
of $\psi$  will follow the $J-$factor and will be centered at the
Galactic centre, where $\cos \psi=\cos l \cos b$. However,
current Fermi-LAT observations suggest that the number of signal
photons is very limited ($N<$ 35 for current ``most
optimistical" signal and usually we have $N<20$). In such a case, due to the limited
statistics, the spatial distribution of the signal photons can be
very different from the ideal morphology, as shown below.

\begin{figure}
\centering
\includegraphics[width=15.0cm, angle=0]{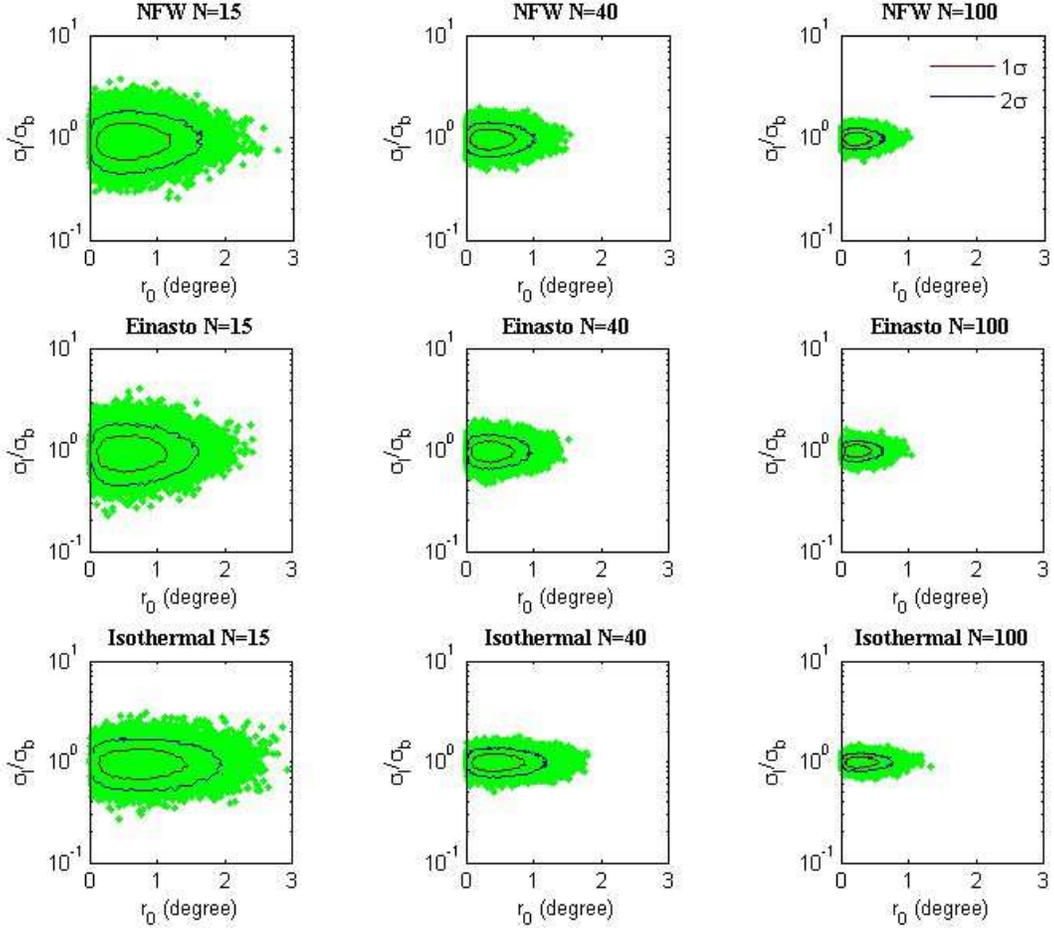}
\caption{ The distribution of photons about the elongation rate
($\sigma_{\ell}/\sigma_b$) and the offset angle from the Galactic
center for the combinations of three dark matter distributions (
NFW, Einasto, and Isothermal) with three numbers of detected photons
($N$=15, 40, 100). } \label{Fig3}
\end{figure}

In the Monte Carlo simulations, following Yang et al. (2012) we assume that the photons are from an angle $\psi\leq 5^\circ$ around
the Galactic center.
We simulate 100,000 observations with $N=15,~40,~100$ photons each.
The average center of the photons is given by
$\ell_0=\sum\limits_{i=1}^{N} \ell_{\rm i}/N$ and $b_0=\sum\limits_{i=1}^{N} b_{\rm i}/N$, and
the offset of the morphology center from the Galactic center is $r_0=\sqrt{\ell_0^{2}+b_0^{2}}$. Following Yang et
al. (2012) we also define the elongation rate
\begin{equation}
\sigma_{\ell}/\sigma_b=\sqrt{\frac{1}{N}\sum\limits_{i=1}^{N}(\ell_{\rm i}-
\ell_0)^{2}} \left/ \sqrt{\frac{1}{N}\sum\limits_{i=1}^{N}(b_{\rm i}-b_0)^{2}}
\right.
\end{equation}
to describe the asymmetric property of the photon map.
In Fig.3 we show the distribution of photons about the
elongation rate ($\sigma_{\ell}/\sigma_b$) and the offset angle from
the Galactic center ($r_0$). Lines in this figure present the $1\sigma$
and $2\sigma$ contours.
The probability of $r_0>1.5^{\circ}$ is about $1.95\%$, $1.38\%$, and
$6.29\%$ for the generalized NFW, Einasto, and Isothermal DM density
profiles in the case of $N=15$, respectively. Such a fact strongly suggests that a sizable offset will be observed. Our prediction will be
directly tested by the ongoing and upcoming high energy
observations.
With the increase of photon statistics, the
deviation of the morphology center from the real center decreases, as expected.
Then the chance of observing a large offset is
accordingly much smaller. For example, in the case of $N=40$ we
have
$P(r_0 >1.5^{\circ})=10^{-5},~9.9\times10^{-6},~6\times10^{-4}$
for the generalized NFW, Einasto, and Isothermal DM density profile,
respectively. That is to say, the morphology will be more symmetric
if more photons are detected.

\begin{figure}
\centering
\includegraphics[width=14.0cm, angle=0]{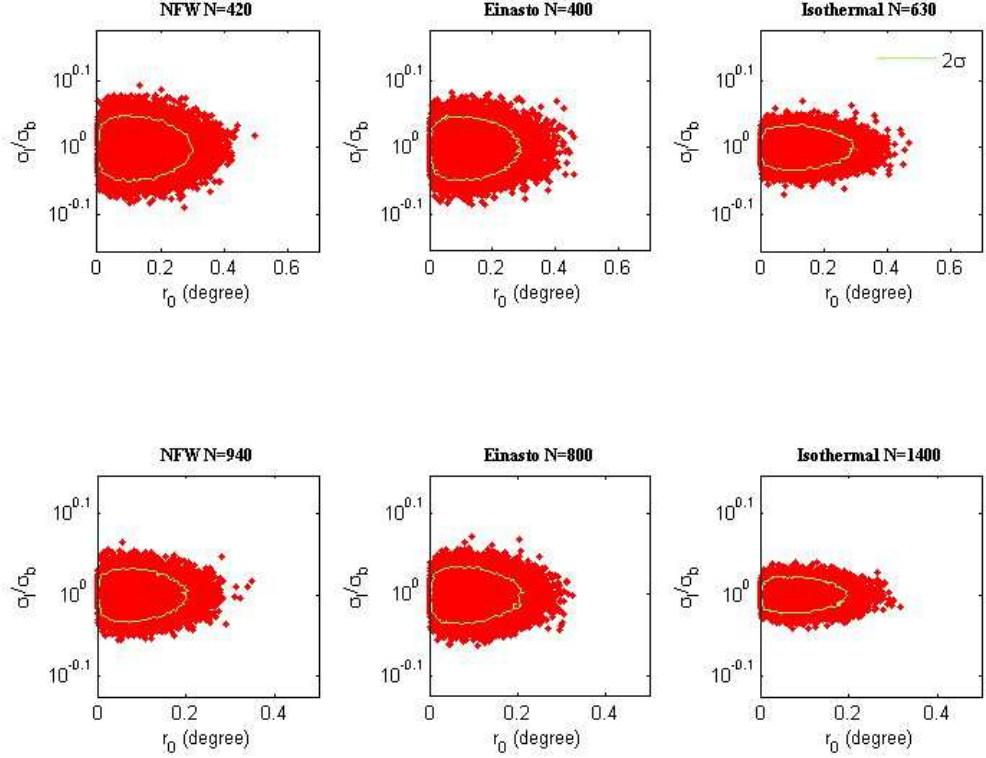}
\caption{The needed $\gamma-$ray line photon number to reliably
constrain whether an offset is due to the limited statistics or alternatively due to the non-standard spatial
distribution of DM particles with $r_{\rm c}\sim 0.3^{\circ}$ and $\sim 0.2^{\circ}$ (top and bottom, respectively) at a confidence level of $2\sigma$.} \label{Fig4}
\end{figure}

So far we have assumed that the distribution of dark matter particles in the
central Galaxy is centered at
$(\ell,~b)=(0^{\circ},~0^{\circ})$. If it is not the case larger offsets
are expected. For example,
if we assume that the dark matter distribution is still
spherically symmetric but centered at $(\ell_{\rm c},~b_{\rm
c})=(1.0^{\circ},~0^{\circ})$, it is straightforwardly to show that
the simulated distribution of the line signal photons would be
centered at $(\ell,~b)=(1.0^{\circ},~0^{\circ})$ and an offset from
the center will be very likely. One possible example for the
non-standard dark matter distribution can be found in Kuhlen et al.
(2013). Therefore, so far at least two effects can give rise to the
offset of the line signal photons. One is mainly due to the limited
statistics. The other is due to the non-standard spatial
distribution of the dark matter particles (i.e., $\ell_{\rm c}\neq
0$ or $b_{\rm c}\neq 0$, or both). In principle, in the future we
can distinguish between these two possibilities. However, the number
of the needed signal photons is expected to be a few hundreds. In
Fig.4 we have presented the needed photon number to
reliably constrain whether an offset is due to the limited
statistics or alternatively due to the non-standard spatial
distribution of DM particles. For example, to exclude the
non-standard Einasto spatial distribution model with $r_{\rm
c}=\sqrt{\ell_{\rm c}^{2}+b_{\rm c}^{2}}=0.3^\circ$ at a confidence
level $\sim 2\sigma$ we need $N\sim 400$. It is not an easy task to
collect so many photons (see Fig.1). The upcoming high energy resolution
detectors such as DAMPE and CALET have a much higher energy
resolution than Fermi-LAT and are thus ideal instruments to identify
line-like $\gamma$-ray signal (e.g., Li \& Yuan 2012). However,
these two detectors have an ``acceptance" (i.e., the Effective Area
times the field of view (FoV)) smaller than Fermi-LAT (see
Table 1), which limits the detectable number of the
$\gamma-$ray line signal. We thus do not expect to get a perfect
coincidence of the signal region with the expected DM distribution
even in future space-based observations. The situation may be
changed for the future ground-based observations. For example with
the Cherenkov Telescope Array (CTA)
(http://www.cta-observatory.org/), the goal of
reliably identification of the dark matter distribution in the very
inner Galaxy may be achieved.

\begin{table}
\caption{Comparison of the performance of Fermi-LAT (Atwood et al.
2009), CALET (Masaki et al. 2013) and DAMPE (J. Chang, 2013, private communication).}
\begin{tabular}{l p{4cm}<{\centering} p{4cm}<{\centering} p{4cm}<{\centering}}
\hline
Parameter & FERMI & CALET & DAMPE\\
\hline
Energy Range  & 20 MeV-300 GeV & 4 GeV-10 TeV& 2 GeV-10 TeV  \\
Effective Area & 7600 cm$^2$  & 600 cm$^2$ & 3600 cm$^2$  \\
Field Of View &  2.5 sr & 2.0 sr &  1.0 sr \\
Energy Resolution  & 10\% (100 GeV) & 2.5\% (100 GeV) & 1.5\% (100 GeV) \\
Angular Resolution  & $ 0.25^\circ$(10 GeV)& $ 0.35^\circ$ (10 GeV) & $0.25^\circ$ (10 GeV) \\
\hline
\end{tabular}
\end{table}

In summary, we have investigated the spatial distribution of the
line signal photons originating from the dark matter annihilation.
Three representative dark matter distribution models have been
adopted and we find out that usually the center of the signal region
will be different from the center of the dark matter distribution
unless the photon number is huge. Even for the future space-based
observations, the expected line signal photons are still too small
to reliably constrain the center of the dark matter distribution.

\begin{acknowledgements}
We thank R. Z. Yang, X. Li, and J. Chang for valuable suggestions
and detailed comments. This work was supported by National Natural
Science of China (Grant No. 10925315, 10973041, and 11303107), and
by the Science and Technology of Development Fund of Macau (Grant
No. 068/2011/A).
\end{acknowledgements}

\end{document}